\documentstyle[twoside,12pt]{article}

\begin{document}
\setlength{\baselineskip}{3.25ex}
\def\maybepagebreak{\vfill\eject}
\parskip 12pt plus 1pt

\newfont{\rmsmall}{cmr12 scaled 900}

\def\degree{\hbox{$^\circ$}}
\def\deg{\hbox{$^\circ$}}

\title{\centerline{CO (J=4$\rightarrow$3) and [C I] Observations}
\centerline{of the Carina Molecular Cloud Complex}}

\bigskip
\bigskip

\author{
Xiaolei Zhang$^{1,2}$ \\
Youngung Lee$^{1,3}$ \\
Alberto Bolatto$^{4}$ \\
Antony A. Stark$^1$ }

\maketitle

\centerline{Accepted for publication in the {\em Astrophysical Journal}}

\bigskip
\bigskip
\bigskip

\noindent 1. Smithsonian Astrophysical Observatory
\newline \noindent 60 Garden Street, Mail Stop 78
\newline \noindent Cambridge, MA 02138

\noindent 2. Present Address: Raytheon ITSS and
\newline \noindent NASA Goddard Space Flight Center
\newline \noindent Code 685
\newline \noindent Greenbelt, MD 20771

\noindent 3. Korea Astronomy Observatory and
\newline \noindent Taeduk Radio Astronomical Observatory
\newline \noindent Whaam-dong, San 36-1, Yusung,
\newline \noindent Taejon 305-348, Korea

\noindent 4. Department of Astronomy and
\newline \noindent Institute for Astrophysical Research
\newline \noindent Boston University
\newline \noindent 725 Commonwealth Avenue
\newline \noindent Boston, MA 02215

\section*{Abstract}

We present large area, fully-sampled maps of the Carina molecular
cloud complex in the CO ($ J = 4 \to 3 $) and neutral carbon
[C~{\rmsmall I}] $^3P_1 \rightarrow ^3P_0$ transitions.  These data
were obtained using the 1.7 meter diameter Antarctic Submillimeter
Telescope and Remote Observatory (AST/RO).  The maps cover an area of
approximately 3 square degrees with a uniform $1'$ spatial sampling.
Analysis of these data, in conjunction with CO ($J = 1 \to 0$) data
from the Columbia CO survey and the IRAS HIRES continuum maps for the
same region, suggests that the spiral density wave shock associated
with the Carina spiral arm may be playing an important role in the
formation and dissociation of the cloud complex, as well as in
maintaining the internal energy balance of the clouds in this region.
Massive stars form at the densest regions of the molecular cloud
complex.  The winds and outflows associated with these stars have a
disrupting effect on the complex and inject mechanical energy into the
parent clouds, while the UV radiation from the young stars also heat
the parent clouds.  The present set of data suggests, however, that
massive stars alone may not account for the energetics of the clouds
in the Carina region.  The details of the data and the correlation
among the various data sets hint at the possible role that the spiral
density wave shock plays in feeding interstellar turbulence and
in heating molecular clouds.

\section{Introduction}

\subsection{Submillimeter-wave Studies of Molecular Clouds}

Studies of the Interstellar Medium (ISM) on large scales
are often pursued through radio surveys because the Galaxy is transparent 
at radio wavelengths and because interstellar gas in its various
phases emits radio lines which can be observed in emission over large
areas of sky.  The line shapes are easily resolved by 
radio spectroscopic techniques and
reflect the motion of the ISM on the galactic scale and
the internal dynamics of clouds. 
The past half-century of progress in radio techniques have seen
dramatic improvements in sensitivity (cf. Carlstrom \&
Zmuidzinas 1995) as well as a general trend towards
higher frequencies and the application of those newly-available 
frequencies to the study of interstellar material.

This paper describes the study of a region of roughly
150$\times$100~pc in size surrounding the bright southern peculiar
star $\eta$ Carinae, observed in the mid-submillimeter lines of CO ($
J = 4 \to 3 $ at 460 GHz) and neutral carbon ([C~{\rmsmall I}] $^3P_1
\rightarrow ^3P_0$ at 492 GHz).  The CO ($ J = 4 \to 3 $) line is a
tracer of the warm ($T\sim 50\, \rm K$) and dense ($n \sim 10^5 \,
{\mathrm cm^{-3}}$) cores in molecular clouds (Viscuso \& Chernoff
1988), whose average properties as seen in CO $ J = 1 \to 0 $ line
studies are colder ($\sim 10 \, \rm K$) and more diffuse ($ n \sim
10^3 \, {\mathrm cm^{-3}}$).  The excitation of the CO ($ J = 4 \to 3
$) transition thus requires high densities, similar to those of common
tracers of dense gas such as ${\mathrm H_2CO}$ (e.g., Magnani, La Rosa
\& Shore 1993) and CS $J = 2 \to 1$ (e.g., Lada, Bally \& Stark
1991). Unlike these, however, it also requires warm temperatures.

The [C~{\rmsmall I}] line, on the other hand, is
expected to trace the photon-dominated regions (PDR) in the outer
envelopes of molecular clouds (Tielens \& Hollenbach 1985). In these
regions neutral carbon is found in a thin layer between C$^+$ and CO,
determined by the equilibrium between photoionization/recombination
processes on the C$^+$/C$^0$ side, and photodissociation/molecule
formation processes on the C$^0$/CO side.  The [C~{\rmsmall I}] $^3P_1
\rightarrow ^3P_0$ transition has a minimum excitation temperature of
24K and critical density $ n \sim 10^3 \, {\mathrm cm^{-3}}$ for
collisions with H$_2$ (e.g. Schroder et al. 1991).  Therefore, it is
easily excited in dense interstellar gas. Observations show that the
[C~{\rmsmall I}] emission is surprisingly well-mixed and
well-correlated with ${\mathrm ^{12}CO}$ and ${\mathrm ^{13}CO}$
emission in the $J = 1 \to 0$ and $J = 2 \to 1$ transitions (e.g., Plume
1995; Keene et al. 1997).  Stutzki et al. (1988) suggest that this
effect results from the clumpiness of molecular material, so that the
``surface'' layers are distributed throughout the volume of the cloud.
Clumpy PDR models (Meixner \& Tielens 1993, 1995; Spaans 1996) 
produce levels of [C~{\rmsmall I}] emission
similar to what is observed near star-forming regions in molecular
cloud cores, but have not been entirely successful in explaining the
surprising uniformity of [C~{\rmsmall I}] in the bulk of molecular
clouds (Keene et al. 1997).

\subsection{Giant Molecular Cloud Complexes}

It is a well known fact that the most massive molecular cloud complexes are
concentrated in spiral arms (Stark 1979; Elmegreen 1979; Dame et
al. 1986).  Indeed, massive stars, H~{\rmsmall II} regions, and dust
lanes, which are the visual tracers of spiral arms, are all
manifestations of concentrations of giant molecular clouds. The
processes leading to the formation of these complexes are not
completely understood, but the observational evidence suggests they
are strongly linked to the passage of the gas through the spiral
density wave shock.  Indeed, the role of spiral density wave shock in
the formation and evolution of the galactic giant molecular cloud
complexes has been speculated upon and investigated since the advent
of the density wave theory, and this effort has continued throughout
the past decades (Roberts 1969; Elmegreen 1979; Balbus \& Cowie 1985;
Dame et al. 1986; Heyer 1998; Zhang 1998).

The problem of cloud complex formation and dissociation 
are closely related to two other issues: 
\begin{enumerate}
\item{ The 
source of the supersonic turbulence energy injection into the clouds
(see, e.g. Larson 1981; Myers 1983 and the references therein).
Turbulence has a natural tendency to cascade downward and dissipate
into heat and line radiation at the smallest scales.  For the galactic
molecular clouds, the time scale for this cascade is on the order of a
free-fall time for the largest clouds (i.e., $\sim 10^6 \, {\mathrm
yr}$), much shorter than the lifetime of molecular clouds (Larson
1981).  Therefore, turbulent energy must be constantly injected into the 
interstellar medium (ISM) to sustain the supersonic linewidths 
observed in galactic
molecular clouds.  The size-linewidth relation connects the physical
size of a region with the observed linewidths, and it is observed to
hold over four orders of magnitudes in cloud size (Larson 1981; Myers
1983).  The small-scale energy injection mechanisms considered (e.g.,
stellar winds and outflows) usually fail to 
either generate sufficient energy
injection or reproduce the correct form of the size-linewidth
relation.  Associated with the issue of the source of turbulence
energy input is the issue of how and where the turbulence energy is
dissipated.}
\item{The processes dominating
the overall energy balance in the ISM.  In recent years the theory of
PDRs (believed to constitute more than 90\% of the galactic ISM) has
gradually confronted serious challenges as observational data
accumulate.  In a recent review article, Hollenbach and Tielens (1999)
cite several instances of observations of Galactic and extragalactic
star-forming regions where the current theory of the PDR often
produces a much a lower temperature than that measured in the
rotational quadrupole transitions of H$_2$, and a much higher ratio of
[C~{\rmsmall II}]$/L_{\mathrm FIR}$ than is seen in infrared luminous
regions.  These authors conclude that these regions must have
additional sources of energy input in order to account for the energy
balance of the PDR.  These sources may include the dissipation of
magnetohydrodynamic turbulence.  }
\end{enumerate}

Recent work on the theory of the dynamics and evolution of spiral
galaxies (Zhang 1996, 1998, 1999) indicates that there is significant
energy and angular momentum exchange between a quasi-steady spiral
density wave and the basic state of the galactic disk.  This exchange
process is of such a magnitude that it should significantly affect
molecular cloud complex formation and dissociation.  An important
consequence of this process is that the orbiting disk matter,
including both stars and gas, receives random-motion energy injection
each time it crosses a spiral wave crest.  The amount of this energy
injection is found to be of the right magnitude to feed the
interstellar turbulence and support the cascade of turbulent energy to
the small scales and subsequent dissipation (Zhang 2000).  Thus, the
spiral density wave may play an important role in the internal energy
balance and the turbulent motions of the galactic molecular clouds. A
clear demonstration of the relation between the spiral density wave
and molecular clouds, however, is yet to be established.  It is in
this context that we have selected the Carina molecular complex as the
region of our study.

\subsection{The Carina Molecular Cloud Complex}

The Carina molecular complex is a segment of the Carina spiral arm
surrounding the extraordinary Luminous Blue Variable (LBV) star $\eta$
Carinae.  It is located between Galactic longitudes $284\deg$ and
$289\deg$ and latitudes $-2\deg$ and $1\deg$.  Figure 1 is a large
area CO ($ 1 \to 0 $) map from the Columbia CO survey of the southern
Milky Way (Grabelsky et al. 1988) showing this complex.

Situated near the center of the Carina molecular complex is the Carina
nebula, which contains an extremely bright and extended OB association
(Car OB1) and a bright H~{\rmsmall II} region, NGC 3372.  In a region
about 40 pc in diameter there are 64 O type stars (including 6 of the
only 11 type O3 stars known in the Milky Way), $\eta$ Carinae itself,
and a Wolf-Rayet star (Walborn 1995).  This high concentration of the
earliest type stars is unique in the Galaxy.  The nearest region of
higher concentration surrounds 30 Doradus region in the Large
Magellanic Cloud. Many spectroscopic and morphological studies of the
Carina region have been made, covering the entire spectral range from
centimeter to X-rays.  A good overview of the physical conditions in
this region can be obtained from the many contributions in the July
1995 issue of Revista Mexicana de Astronomia y Astrofisica, entitled
``The $\eta$ Carina Region: A Laboratory of Stellar Evolution''.

Past study of the Carina region has focused mainly on the peculiar
star $\eta$ Carinae and on the H~{\rmsmall II} region surrounding it.
The larger molecular complex is mapped by the Columbia CO survey with
a $8.8'$ beam, and by the IRAS satellite in its four spectral bands.
These are large surveys which are not particularly focused on the
Carina region.  We have chosen to map the entire complex with the
AST/RO telescope.  We intended to use the AST/RO data, combined with the
existing survey data, to study the large-scale physical conditions in
this region, and investigate the role of spiral density wave in the
formation and dissociation of molecular cloud complexes.  AST/RO is
very well-suited for this work, because it was designed as a Galactic
survey instrument.

The Carina complex has a very clear line of sight, with a mean color
excess ${\mathrm E_{B-V} = 0.5}$ at a distance of 2.5 kpc (Feinstein
1995).  The various clouds and sub-complexes are distributed along the
Galactic plane in what appears to be a sequential order. Their
kinematics suggest that locations of decreasing longitude correspond
to advancement in the spiral-arm-crossing phase, as can be seen
for those cloud clumps nearest to the Sun on the well-delineated
Carina spiral arm in Figure 4a of Grabelsky et al. (1988).
This correspondence is further supported by cloud morphology (see
Figure 2): clumps near the nebula appear to be coherently shocked,
while clouds to the north are more fragmented.  Moreover, there is
an age gradient in the various star clusters across the complex.  The
Tr 14 and Tr 16 clusters within the Carina nebula (NGC 3372) at
$(l, b)=(287.6\deg$, $-0.65\deg$) are the youngest (age $\sim 10^6 \, \rm{yr}
$) and IC 2581/NGC 3293 at ($284.7\deg$, $0.1\deg$) are the oldest
(age $\sim 5 \times 10^6 \, \rm{yr}$).  They are separated by a
projected distance of $\sim \, 130 {\mathrm pc}$.  This sequential
arrangement is advantageous to the study of the evolution of physical
conditions in the clouds, as the different clouds stream across the
spiral arm.

The Carina nebula itself contains an archetypical outflow (Duschl et
al. 1995) centered on $\eta$ Carinae at ($287.6\deg$, $-0.64\deg$), and
the highest concentration of early type stars known in the Galaxy in
the two ionizing clusters Tr 16 (which is centered on $\eta$ Carinae, and
which also includes a smaller cluster Cr 228 to the south) and Tr 14
(about $10'$ to the north of $\eta$ Carinae).  The region offers the
opportunity to study and possibly disentangle the effects of energy
input to the molecular clouds by massive stars and by spiral density
wave shocks.

In \S 2 and 3, we describe the observational results
and analysis of the Carina region. Sensitive receivers and the clear skies
of the South Pole have permitted extensive mapping of the
CO ($J = 4 \to 3$) and [C~{\rmsmall I}] lines over a region 
including several molecular clouds and covering a segment 
of the Carina spiral arm.  These maps are less biased to
cloud cores and known heating sources than was previously possible,
and therefore allow us to investigate the important question of the relation
between molecular clouds and the environment where these clouds
form and dissociate.

\section{Observations and Data Reduction}

The CO ($4 \to 3$) and [C~{\rmsmall I}] data presented here were
obtained during the Austral winter of 1998, using the 1.7~m telescope
of the Antarctic Submillimeter Telescope and Remote Observatory (Stark
et al. 1997; Lane \& Stark 1996) located at the United States
Amundsen-Scott South Pole Station.

The CO ($4 \to 3$) data were taken during the Austral fall and have
system temperatures between 1500 K and 3000 K. The [C~{\rmsmall I}]
data set was acquired during the months of July through September,
with the system temperature ranging from 1200K to 2200K. Both maps
were obtained sampling on a $1'$ grid, with an integration time of 60
s per point at most locations.  Half of the integration time was spent
on source and half on the two reference positions, situated $\pm90'$
away from the mapping center in RA (which is the same as in Az for a
telescope located at the geographic South Pole).  The reference
positions are free of emission in the Columbia survey map.  The entire
data set was acquired in less than three weeks.

The line strength is calibrated using warm and cold loads, together
with the sky measurement at a location near the source.  Skydips are
done roughly twice a day to assure the stability of the telescope
efficiency and to check the consistency of the single-slab atmospheric
model used to correct for atmospheric absorption.  The beam size for
the SIS quasi-optical receiver used for [C~{\rmsmall I}] observations
was $\sim3.5'$, and the beam size for the SIS waveguide receiver used
for CO ($ 4 \to 3 $) observations was $\sim3'$. These sizes are
estimated by scanning the beam across the limb of the full Moon.  The
main beam telescope efficiency ($\eta_{mb}$) for both receivers is
$\sim70$\%, as estimated from the skydip measurements.  The backend
used was the 2048 channel acousto-optical spectrometer, with a
spectral resolution of 0.4 ${\mathrm km \, s^{-1}} $ (Schieder, Tolls
\& Winnewisser 1989).  AST/RO's pointing was carefully monitored using
the source G291.28-0.72 (located at a declination similar to the
Carina nebula), which was observed every 8 hours (Huang et al. 1999).
The rms pointing accuracy for both data sets is estimated to be better
than $30''$.  The calibrated data is expressed as $T_{\mathit R}^{*}$
(Kutner \& Ulich 1978), which is essentially the same as $T_A^{*}$ for
AST/RO.

The raw data were corrected for atmospheric absorption and a linear
baseline was removed using the software package COMB.  The data cube
thus generated has been further analyzed using the software packages
IRAF and AIPS and plotted using the software packages PGPLOT and WIP.

\section{Results}

In the this section we will introduce the submillimeter data
and discuss three of its properties: 
\begin{enumerate}
\item {the extent of the CO ($4 \to 3$)
and [C~{\rmsmall I}] emission,} 
\item{the spatial and velocity correlation 
between both submillimeter transitions, 
and} 
\item{their relation to the
CO (J=$1 \to 0$) emission.}
\end{enumerate}
 
Figure 3 shows an overlay map of the AST/RO [C~{\rmsmall I}] and CO
($4 \to 3$) observations. The maps were obtained by integrating the
calibrated data cube over the entire velocity range of the Carina
complex, that is, between $-50 \, {\mathrm km \, s^{-1}}$ and $-9
\,{\mathrm km \, s^{-1}}$.  This velocity range is chosen to coincide
with that used by Grabelsky et al. (1988) for the CO ($1 \to 0$) data,
although there appears to be small amount of emission belonging to the
complex at velocities as high as $0 \,{\mathrm km \, s^{-1}}$.  Each
map is a composite of 3 individual, partially overlaping, submaps for
each transition.  While the southernmost and northernmost submaps have
identical size in [C~{\rmsmall I}] and CO ($4 \to 3$), the central
submap in [C~{\rmsmall I}] is significantly smaller than its CO ($4
\to 3$) counterpart, as shown by the color and contour boundaries in
Figure 3.

From Figure 3, it is immediately evident that the CO ($4 \to 3$) and
[C~{\rmsmall I}] transitions are approximately coextensive throughout
the whole Carina region.  This is the first instance in which a higher
transition of CO and the [C~{\rmsmall I}] emission are found to be
coextensive over such a large area, spanning approximately
$150\times100$ parsecs.  Previous studies have found a similar result
for the lower transitions of CO as well as $^{13}$CO with [C~{\rmsmall
I}] (e.g., Philipps \& Huggins 1981; Keene et al. 1985, 1997; Plume
1995, and the references therein).  This result is noteworthy, since
CO ($4 \to 3$) requires warm, high density conditions to be excited.

The extent of the [C~{\rmsmall I}] emission over the entire region is
also remarkable.  The theoretical expectation is that [C~{\rmsmall I}]
arises as a result of the photodissociation of CO in the PDR occurring
at extinction $A_{\rm v}<3$ (Keene et al. 1997).  Usually the
coextensiveness of the [C~{\rmsmall I}] and CO emission is attributed
to the fact that the ISM is clumpy, and therefore porous to UV
radiation.  We see copious [C~{\rmsmall I}] emission arising from
clouds located far away from UV sources, for example those in Region 6
(Figure 2). Recent PDR modeling results by Kaufman et al. (1999)
indicate that the intensity of the [C~{\rmsmall I}] transition is
insensitive to the radiation field.  The nearby open clusters
(NGC~3293 and NGC~3324) possess one type O star and several early type
B stars (Clari\'a 1977; Feinstein \& Marraco 1980), and may thus be
capable of photodissociating CO several parsecs away.

The intensity peaks of the [C~{\rmsmall I}] and CO ($4 \to 3$)
distributions are approximately coincident, and are located near the
infrared peak illuminated by the compact star cluster Tr 14 (Figure
12).  Small differences in the morphology of the two transitions do
exist, however. Most noticeable is the double peaked structure of
Region 5 in [C~{\rmsmall I}], which exhibits only one peak in CO ($4
\to 3$).  The second peak of neutral carbon is probably associated
with an embedded source that must have photodissociated most of the
surrounding CO (Figure 10).  Figures 4 and 5 show the CO ($4 \to 3$)
data overlayed on the lower resolution Columbia CO (J=$1 \to 0$) map
($8.8'$ resolution, Grabelsky et al. 1988) and on the high-resolution
MOPRA map ($1.4'$ resolution, Brooks et al. 1998).

A closer look at the velocity information reveals that both lines
display identical kinematics.  Figures 6 and 7 show velocity channel
maps for the [C~{\rmsmall I}] and CO ($4 \to 3$) transitions.  These
are essentially identical in both species, indicating that the
[C~{\rmsmall I}] and CO ($4 \to 3$) emitting gas are well mixed.
%Most of the emission occupies the velocity
%range from $-40 \, {\mathrm km \, s^{-1}}$ to $-5 \, 
%{\mathrm km \, s^{-1}}$, although at the top right
%region of the map a small clump, which
%also appears in the Columbia CO ($1 \to 0$) map,
%extends to  $0 \,{\mathrm km \, s^{-1}}$. 
A few clumps, most noticeably in Region 3, do not quite follow the
general trend determined by Galactic rotation. This is perhaps an
indication that they are perturbed by the activity (i.e., winds and
outflows) surrounding the star $\eta$ Carinae. 

Figure 8 displays the composite spectra for Region 1 through 6, where
the similarity of the line profiles can be appreciated.  The strongest
emission arises from Region 3, where there is a clearly non-gaussian
line profile. This is the signature of the gas entrained in the
bipolar outflow from $\eta$ Carinae.  The double peaks observed for
region 2 are also likely to be produced by the interaction of the
ambient gas with the winds and outflows originating in $\eta$ Carinae.

\section{Discussion}

\subsection{Physical Conditions in the Region}

A comparison of the [C~{\rmsmall I}] and CO ($4 \to 3$) distributions
as shown in Figure 3 gives us some qualitative idea of the temperature
and density distribution in the mapped region, after taking into
account the excitation conditions for these two species.  Estimates of
the physical conditions of the gas independent of chemistry can be
obtained by using two transitions of the same chemical species. The CO
($4\to 3$)/($1\to 0$) ratio is sensitive to a combination of density
and temperature, but mostly to density when $T>50$ K and, and mostly
to temperature when $n>10^5$ cm$^{-3}$ and $T<50$ K.  While
temperatures $T>50$ K are very rare in molecular clouds over large
spatial scales, densities $n\sim10^5$ cm$^{-3}$ necessary to
thermalize the CO ($4\to 3$) transition are more commonly
found. Furthermore, in opaque cores radiative trapping easily reduces
this density requirement to $n\sim10^4$ cm$^{-3}$, which is a typical
molecular cloud density over large spatial scales. Thus, although the
CO ($4\to 3$)/($1\to 0$) intensity ratio is affected by both density
and temperature, in the discussion following we will assume that the
prevalent effects are due to temperature.

A quantitative estimate of the average excitation temperature for a
region can be obtained assuming thermalized but optically thin
emission. After convolving our CO ($4\to 3$) data to the resolution of
the Columbia CO ($1\to 0$) survey for this region (generously provided
by T. Dame), a rough estimate of the gas temperature is given by:
\begin{equation}
T_{\mathrm{ex}} \sim {{55 \, \mathrm{K}}\over{\mathrm{ln} \left(
 {{16 T_{1\to 0}^{peak}} \over { T_{4\to 3}^{peak}}}  \right)}} ~~ .
\label{eq:1}
\end{equation}

Note that use of this equation assumes both the local thermodynamic
equilibrium (LTE) and optically thin conditions.  While LTE is
expected to hold in dense molecular clouds, the optically-thin
emission certainly does not hold for CO.  Here we assume that the
clouds are composed of many smaller optically-thick cloudlets (i.e.,
the ``mist model'', Dickman et al.  1986; Solomon et al. 1987).  We
further assume that the fluxes we measured in the two transitions are
proportional to that emitted by the optically-thin envelopes of the
constituent cloudlets.  Thus in taking the ratio of the fluxes in
these two transitions, we are effectively using the optically thin
sub-components of the original cloud to estimate its
temperature.
This procedure has been found to give consistent 
temperature estimates when using two pairs of adjacent CO transitions.
For example, the excitation temperature obtained from the CO ($2 \to
1$) and CO ($1 \to 0$) ratios has been found to be very close to the
excitation temperatures obtained from the CO ($3 \to 2$) and CO ($2
\to 1$) ratios for the same regions (Zhang 1992). Thus, we are confident
that despite the caveats Eq. \ref{eq:1} can be used to obtain a rough
estimate of the physical temperatures.

Figure 9 shows the excitation temperature map derived using
Eq. \ref{eq:1} and the ratio of integrated intensities (i.e.,
implicitly assuming similar linewidths for both CO transitions).  This
method will produce spuriously high temperatures if the CO ($1 \to 0$)
emission is self-absorbed, but fortunately there is
little evidence for such problems in this map.  Apart from the noise
near the boundaries of the emitting regions, it is apparent that the
highest average excitation temperature is obtained in the vicinity of
$\eta$ Carinae.
%On very small scales near the star, the temperature of the 
%stellar atmosphere of $\eta$ Carinae is over 8000K as
%estimated using Fe II lines (Thackeray 1967;
%Viotti 1969).  
There is also an excitation temperature gradient across the entire
map, with higher temperatures in the southern regions.  This is likely
due to the combined effect of two energy inputs: 1) the southern region
is situated near the ionizing star cluster, and 2) it is near the spiral
density wave crest, experiencing energy injection during arm crossing.
Naturally, these two sources of energy are difficult to separate.  In
the six regions marked in Figure 2, the average excitation
temperatures are respectively, $\rm T_{ex1} = 34\,
\rm K$, $\rm T_{ex2}=52 \, \rm K$, $\rm T_{ex3}= 33\, \rm K$, $\rm
T_{ex4}= 24 \,
\rm K$, $\rm T_{ex5}= 19 \, \rm K$, and $\rm T_{ex6}=10\, \rm K$.  
By comparison, Ghosh et al. (1988) found a dust temperature $\sim40$ K
in the nebula region (our Region 2). The similarity of these two
temperatures suggests that our estimates based on Eq. \ref{eq:1} are
not unreasonable.

While these average excitation temperatures are lower than the nominal
excitation threshold for the CO ($4 \to 3$) transition of 55 K, they
presumably represent the mean excitation temperatures over an ensemble of
clumps within each region.  From the excitation temperatures obtained
above, we see that these regions have a much hotter temperature than
that expected for well-shielded, dark clouds (i.e., 8---10 K).  This
is true even for regions 4 and 5 located farther away
from the Carina nebula and star clusters.

The far-infrared (FIR) maps of the Carina complex reveal the position
of the heating sources, and their association with the molecular
peaks.  We obtained IRAS HIRES (resolution-enhanced) maps of this
region in all 4 bands (12 $\mu$m, 25 $\mu$m, 60 $\mu$m, and 100
$\mu$m). Figures 11 and 12 show the overlay of the CO ($4 \to 3$)
emission with the IRAS HIRES 12 $\mu$m and 100 $\mu$m maps,
respectively. Figure 10 shows the overlay of our [C~{\rmsmall I}] map
with the IRAS HIRES 100 $\mu$m continuum.  The FIR peaks are
concentrated near the Carina nebula region, with the brightest
emission peak at 100 $\mu$m closely associated with the main CO peak
in Region 3.  For shorter wavelengths, the FIR emission peak shifts
gradually towards the ionizing cluster Tr 14, as discussed by Cox
(1995).  The region near the star $\eta$ Carinae (the bright far-IR
point source at $l = 287.6\deg$, $b = -0.64\deg$, between the two
molecular clumps) appears free of submillimeter emission, possibly due
to the cavity blown open by the outflow and winds from the star (Cox
1995).  Away from the nebula region, there is a drastic decrease in
the FIR luminosity accompanied by a diminishing number of 12 $\mu$m or
100 $\mu$m peaks.  This suggests that these molecular clouds contain
few embedded young stellar objects. It is unclear whether the
star-formation activity near Regions 4 and 5 is enough to maintain
their cloud temperatures ($T\sim20$ K), or additional energy inputs
are needed.

Is there evidence for unaccounted sources of energy input
in this region? The total FIR luminosity of the nebula region can be
estimated using (Lonsdale et al. 1985, Lee et al. 1996)

\begin{equation}
L_{FIR} = 0.394\, {{\mathrm{L}}_{\odot}} \cdot R(\overline{T_d},\beta) 
\left[{{S_{100} + 2.58S_{60} }\over{1 {\mathrm{MJy \, sr}^{-1}}}}\right] 
\left[{{D}\over{\mathrm{kpc}}}\right]^2 ~~,
\end{equation}

\noindent where $D$ is distance, and the correction 
factor $R(\overline{T_d}, \beta)$ is given by

\begin{equation}
R(\overline{T_d},\beta) = 
(\int_{x_1}^{x_2} {{x^{3+\beta}}
\over
{e^x-1}} dx) /
(\int_{x_3}^{x_4} {{x^{3+\beta}}
\over
{e^x-1}} dx) ~~,
\end{equation} 

\noindent where $x_n \equiv hc/\lambda_n k \overline{T_d}$, 
$\lambda_1=1 \mu$m, $\lambda_2=500 \mu$m, $\lambda_3=42.5 \mu$m,
$\lambda_4=122.5 \mu$m, and we assume $\beta = 1$.  Using an average
dust temperature of $40\,\mathrm{K}$ for the nebula region (Ghosh et
al. 1988), we obtain a total FIR luminosity $L_{FIR}\sim10^7 ~ \rm
L_{\odot}$.  This number is comparable to the total luminosity of all
the OB stars in the nebula,
$L_{stars}\sim2 \times 10^7 \mathrm{L}_{\odot}$ (Feinstein 1969;
Walborn 1973).  Since the nebula region is already a blown-open
cavity, however, we expect that only a small part of the UV flux of
the OB stars will be intercepted by the dust and gas.  Even though the
UV power from stars and the FIR luminosity are comparable, it seems
likely that there is additional energy input to the region,
for example in the form of mechanical energy from shocks.
The extent of the CO ($4 \to 3$) emission in the broader
surrounding area also reinforces the above evidence from
analysing the far-IR emission near the nebula that the
heating of the Carina molecular complex is likely to be contributed
by energy sources in additional to stellar luminosity. 

\subsection{Origins of the [C~{\rmsmall I}] Emission}

A striking feature of this dataset is the large spatial extent of 
the [C~{\rmsmall I}] emission, as well as the
homogeneity of its intensity, both near and far from the UV sources.

We first estimate the column density of neutral carbon using the 
equation

\begin{equation}
N_{\rm C^0} = 1.7 \times 10^{16} \, \mathrm{cm}^{-2}
\left[{{\int T_a^* {\mathrm [C~{\rmsmall I}]} \mathrm{d} v}
\over{1 \mathrm{K \, km \, s}^{-1}}}\right]
\end{equation}

\noindent (Plume 1995). From the observed integrated intensity
of [C~{\rmsmall I}] we obtain a peak C$^0$ column density $N(\rm
C^0)\sim1.4 \times 10^{17}\, {\mathrm cm^{-2}}$. This can be compared
to $N(\rm C^0)\sim1.2 \times 10^{17}\, {\mathrm cm^{-2}}$ measured for
the bulk of the S~140 molecular cloud (Plume 1995), and $N(\rm
C^0)\sim2$--$3 \times 10^{17} \, {\mathrm cm^{-2}}$ as given by the
PDR models (van Dishoeck \& Black 1988; Hollenbach, Takahashi,
\& Tielens 1991).  The brightness of [C~{\rmsmall I}] emission in the Carina 
molecular complex is therefore unexceptional.

According to the 100$\mu$m and [C~{\rmsmall I}] overlay in Figure 10,
it appears that most of the [C~{\rmsmall I}] emission in Regions 2 and
3 is likely to have originated from the photodissociated CO by the
intense UV radiation near the core of the nebula.  The connection of
the [C~{\rmsmall I}] emission in Region 3 to the UV sources is also
manifested by the bar-like profile of the strongest [C~{\rmsmall I}]
emission peak bending towards the ionizing cluster Tr 14 and the
peculiar star $\eta$ Carinae.

In Regions 4 and 5, however, the [C~{\rmsmall I}] emission reaches
almost the same intensity as the peak of Region 3, near the nebula.
In fact, the [C~{\rmsmall I}] emission across the entire molecular
complex ($\sim150$ pc) is extremely homogeneous and appears to have
little correlation with the presence of FIR peaks or the location of
nearby sources of ionizing radiation, aside from some localized
examples. The UV flux $G_0$, however, varies by several orders
of magnitudes as indicated by the FIR intensity.  The average
UV field in the 6 regions can be estimated using

\begin{equation}
G_0 \sim {1\over{1.6\times10^{-3}}}\,{L_{FIR}\over {8 \pi d^2}} ~~, 
\end{equation}

\noindent where $G_0=1$ 
is the interstellar UV radiation field in the vicinity of the Sun ($1.6 \times
10^{-3} \mathrm{erg\, s^{-1} \, cm^{-2}}$; Habing 1967), $d$ is the
average radius of a region, and we are implicitly assuming that all
the UV photons are collected by interstellar dust grains and
reradiated in the FIR.  The $G_0$ values for regions 1 through 6
are approximately 800, 1700, 1000,
170, 130, and $\leq10$ respectively.  The UV
field around the emission peaks can be several order of magnitude
higher than the region average.  The insensitivity of the [C~{\rmsmall
I}] emission to the radiation field incident on the clouds has been
predicted by PDR models (Kaufman et al. 1999) and may provide
the explanation for the observed homogeneities of [C~{\rmsmall I}] emission,
although we do not consider that a unique connection 
between the full content of [C~{\rmsmall I}] in this region
and a photodissociation process is firmly established.

What is the I$_{[C~{\rmsmall I}]}$/I$_{CO(4 \to 3)}$ throughout the
region, and what does it tell us about the density of the molecular
gas according to the standard theories of PDR?  
In Figure 13 we present the scatter plots of the [C~{\rmsmall
I}] and CO ($4 \to 3$) integrated intensities 
(each represented in units of K~km~s$^{-1}$) for regions 1 through 6.
The two transitions are well correlated within each region, whereas
between the different regions the line intensity ratio changes:
I$_{[C~{\rmsmall I}]}$/I$_{CO(4 \to 3)}$={0.21, 0.17, 0.19, 0.32,
0.45, 0.34} for regions 1 through 6 respectively. This roughly
corresponds to a monotonically increasing ratio of I$_{[C~{\rmsmall
I}]}$/I$_{CO(4\to 3)}$ with decreasing Galactic longitude (i.e., 
advancing spiral arm crossing phase, see Fig. 2), due
mostly to the decrease in CO ($ 4 \to 3 $) intensity away from the
nebula region in comparison with the relatively homogeneous
[C~{\rmsmall I}] (Figure 3). Figure 14 shows the I$_{[C~{\rmsmall
I}]}$/I$_{CO(4\to 3)}$ ratio predicted by the standard PDR
calculations (Kaufman et al. 1999).  Using the values of $G_0$ found
in the previous paragraph, and the I$_{[C~{\rmsmall I}]}$/I$_{CO(4 \to
3)}$ ratio measured for the same regions, we can place them on this
plot.  We see that Regions 1, 2, and 3 have an average density
$n\sim10^5$ cm$^{-3}$, while for Regions 4 and 5 we predict somewhat
lower densities $n\sim3\times10^4$ cm$^{-3}$, and the density
prediction is uncertain for Region 6 because the [C~{\rmsmall I}]
intensity there is below the noise level.

Although most physico-chemical models of molecular clouds find that
neutral carbon is predominatly produced by photodissociation of CO and
recombination of C$^+$, some models predict a large fraction of the
gas phase carbon to be C$^0$ ($N$(C$^0$)/$N$(CO)$\sim0.1$ to 0.2)  
at densities below $\sim$ $5.5 \times 10^3 \, {\mathrm cm^{-3}}$
(Pineau des F\^{o}rets et al. 1992; Le Bourlot et al. 1993; 
Flower et al. 1994). The observed correlation between [C~{\rmsmall I}] 
and CO ($4 \to 3$) together with the critical density of
the CO transition ($>10^5 \, {\mathrm cm^{-3}}$) suggests, however,
that this mechanism is not the source of a significant fraction of 
neutral carbon in this region.

\subsection{Origins of the Size-Linewidth Relation}  

In Figure 15 we show the size-linewidth correlation plot for the
molecular clumps derived from the CO ($ 4 \to 3 $) data cube, 
using the clump-finding algorithm developed by one of the
authors (Y. Lee).  The boundary of the clumps is defined to be three
times the rms noise level of the data cube pixels.  The fitted
size-linewidth relation has a slope of 0.6, similar to that found in
other studies of the galactic molecular clouds (cf. Myers 1983 and the
references therein).  Other statistics of the clumps are given in
Table 2.

The major trend of the correlation in Figure 15 is a single linear
relation across the entire complex, regardless of whether a
particular clump lies near or farther away from the $\eta$ Carinae 
outflow. The role of this outflow appears to be 
mainly in perturbing the velocities of several clumps 
(such as the clump in region 2 seen in the velocity
channel maps).  The outflow also perturbs the size-linewidth relation
from a perfect linear correlation---i.e., it adds noise into the
relation.  In fact, the two extreme outliers on Figure 15 are clumps
from Regions 2 and 3, which are most affected by the outflow.  In view
of the basic uniformity of the correlation law across the whole
complex, we conclude that the outflow is not the cause of the
size-linewidth correlation, but is rather a cause for departure from
a perfectly linear relation.

We are, therefore, still in need of a mechanism which is capable of
injecting energy into the interstellar clouds on spatial scales of
hundreds of parsecs.  For the particular region of the Carina
molecular complex at least, many of the proposed large scale
processes, such as supernovae and superbubbles (Kornreich \& Scalo
2000) do not appear to be applicable.  Another proposed mechanism
operating on the galactic level is the coupling of galactic rotational
energy (von Weizsacker 1951; Fleck 1981). However, detailed numerical
simulations have already shown that it is in fact rather difficult to
couple this energy into the internal motion energy of the cloud (Das
and Jog 1995).
 
An alternative candidate mechanism is the spiral density wave. It has
been shown recently that spiral density waves constantly inject energy
into the interstellar medium during spiral arm crossings, at size
scales from 1 kpc down to a few parsecs (Zhang 2000).  Since this
energy is injected through the mediation of the gravitational
potential, it happens simultaneously on large and small spatial
scales.  Using average Galactic spiral parameters, the orbit-averaged
rate of energy injection per unit mass due to the interaction of a
spiral density wave with the disk matter is calculated to be (Zhang
2000)

\begin{equation}
{ {d \Delta E} \over {dt}} 
= 3 \times 10^{-7} \mathrm{(km \, s^{-1})^2 \, yr^{-1}}
.
\end{equation}

Using an average line-width of $\Delta v = 2 ~ \mathrm{km \, s^{-1}}$ 
at size scale of $10 \, \mathrm{pc}$
from Figure 15, the rate of the energy cascade can be found from
the following equation

\begin{equation}
{{\Delta v^3}
\over {L}}
= { { (2 \, \mathrm{km \, s^{-1}})^3} \over { 10 \, \mathrm{pc}} }
= 8.1 \times 10^{-7} \mathrm{(km \, s^{-1})^2\, yr^{-1}}
.
\end{equation}
These two numbers are quite comparable, especially considering that
the energy injection rate during the period of spiral arm crossing is
several times larger than its value averaged over the entire orbital
period.  Energy injection due to the spiral density wave is therefore
a plausible source for maintaining the degree of turbulent motion and
producing the basic trend of size-linewidth correlation observed in
this region.

\section{Conclusions}

We have observed the Carina molecular cloud complex in the CO ($ 4 \to
3 $) and [C~{\rmsmall I}] $^3P_1 \rightarrow ^3P_0$ transitions using
the AST/RO telescope.  We find that throughout the mapped area ($\sim
150\times100$ pc in extent) the CO ($1 \to 0$), CO ($ 4 \to 3 $) and
[C~{\rmsmall I}] emissions are ubiquitous and approximately
coextensive.  The extent and intensity of the [C~{\rmsmall I}]
emission is almost uncorrelated with the location and strength of of
the UV sources. We also find that the clouds in this region appear to 
be warmer than typical dark molecular clouds.

We find that there is a unique size-linewidth correlation
throughout the $\sim150\times100$ pc region, which does not seem to be
related to the spatially confined outflow originating in $\eta$ Carinae.
This suggests that the dominant energy injection mechanism responsible for
turbulence in molecular clouds operates on very large spatial scales,
and is different from localized stellar outflows.

We propose that the same large-scale mechanism could be the energy
source for feeding the interstellar turbulence (thus producing the
observed size-linewidth relation), and increasing the temperature of
these clouds.  We suggest that the spiral density wave shock may play
an important role in the formation and evolution of the molecular
cloud complexes as well as in the energy balance of the clouds. In
particular, the energy injection from the spiral density wave is found
to be of the correct order to produce the observed size-linewidth
relation for molecular clouds, so it might also be responsible for a
part of the heating of the clouds through the dissipation of turbulent
energy.

\section*{Acknowledgments}

We thank Tom Dame and collaborators for making available the Columbia
CO ($ 1 \to 0 $) survey data used in part of the analysis in this
paper, Kate Brooks and collaborators for making available the high
resolution CO ($ 1 \to 0 $) MOPRA data used for comparison, and Mark
Wolfire for making available the results of the PDR calculations.  We
thank the staff at the IPAC in Caltech for processing the IRAS HIRES
maps for this region.  Discussions with and helpful comments from
M. Crosas, T. Dame, M. Heyer, J. Ingalls, G. Melnick and P. Myers are
gratefully acknowledged.  This research was supported in part by the
National Science Foundation under a cooperative agreement with the
Center for Astrophysical Research in Antarctica (CARA) grant DPP
89-20223.  CARA is a National Science Foundation Science and
Technology Center.

\vfill
\eject
 
\begin{verbatim}

Table 1 Known OB associations within the mapping region

-----------------------------------------------------------------------

 name       l        b     D      V     Ntot    Dl     Db    N_KM
          (deg)    (deg) (kpc)  (km/s)         (pc)   (pc)

CAR 1  B  285.98   0.40  2.14   -2.7     24    44.8   74.6    4
CAR 1  C  286.30  -0.16  2.59             8    22.6   31.7    1
CAR 1  E  287.61  -0.68  2.64  -11.0     77    83.0   60.0    3

-----------------------------------------------------------------------

 name:  Association name
 l:     Galactic longitude
 b:     Galactic latitude
 D:     Distance
 V:     Average radial velocity of the association relative to the Sun
 Ntot:  Total number of stars in the association
 Dl:    Size of association along longitude
 Db:    Size of association along latitude
 N_KM:  Number of K and M supergiants

----------------------------------------------------------------------

source: Melnik A.M. & Efremov Yu.N. (1995)

\end{verbatim}

\pagebreak

\begin{verbatim}

Table 2. Cloud clump statistics from the CO 4-3 data set

----------------------------------------------------------------------

    l         b       v        dv     size     Tpk       I_CO    
(degree)   (degree) (km/s)   (km/s)   (pc)     (K)      (Kkm/s)

 287.68    -0.75    -24.38    0.80    1.89     3.06      148.2   
 287.74    -0.62    -23.24    1.10    2.56     3.20      281.2  
 287.38    -0.67    -22.10    1.08    1.37     3.01       73.5 
 286.36    -0.27    -20.50    0.67    1.62     2.80       66.4   
 287.32    -0.56    -16.43    1.40    5.43     5.32     1713.5  
 286.08     0.21    -18.63    1.25    2.53     4.13      390.3 
 288.06    -1.11    -17.43    1.71    7.52     4.05     1273.2  
 287.00    -0.36    -17.85    0.75    2.91     4.52      372.6  
 287.13    -0.54    -17.40    0.53    3.56     3.59      198.4  
 287.13    -0.85    -16.68    0.92    2.74     3.18      225.0  
 287.23    -0.22    -16.34    0.74    2.25     3.26      122.6  
 287.10    -0.72    -16.45    0.58    1.20     2.85       42.7   
 287.25    -0.91    -12.56    0.50    1.16     2.93       27.1  
 285.27    -0.01      3.53    0.62    1.47     3.69       79.7 

\end{verbatim}

\section*{Figure Captions}

\noindent 
Figure 1. Columbia CO survey integrated intensity map
(in units of ${\mathrm K \, km \, s^{-1}}$)
of the fourth quardrant containing the Carina molecular complex
region.

\noindent 
Figure 2. Carina molecular complex region, shown as a contour
plot of CO ($1 \to 0$) of the Columbia data superimposed on the
Digital Sky Survey image (obtained from the SkyView database
of the Goddard Space Flight Center).  The contour level are
10\%-90\%\ of 70 ${\mathrm K \, km \, s^{-1}}$. The three star symbols 
represent the known OB associations in this region, and their
detailed properties are given in Table 1.  The three
concentrations of stars and nebulosities are respectively the
Carina nebula/H~{\rmsmall II} region (NGC 3372), 
NGC 3324 at ($286.2\deg$, $-0.2\deg$),
as well as NGC 3293 and its companion H$_{\alpha}$
region G30 at ($285.9\deg$, $0.1\deg$).  The 6 dashed boxes mark
regions identified for subsquent analysis.

\noindent
Figure 3. AST/RO CO ($4 \to 3$) and [C~{\rmsmall I}] maps.
The contour levels for CO ($4 \to 3$) are 10\%
to 90\% of 80 ${\mathrm K \, km \, s^{-1}}$.  
The half tone for [C~{\rmsmall I}]
is from 0 to 12 ${\mathrm K \, km \, s^{-1}}$.
The extent of the mapping area for each transition is
indicated by the boundaries of the contour or halftone
images.
 
\noindent
Figure 4.  AST/RO CO ($4 \to 3$) map overlayed on the Columbia
CO (1-0) map of the same region.  Note that the AST/RO
map only cover a limited area within that mapped by the Columbia
Survey, as indicated in the previous figure.

\noindent
Figure 5.  AST/RO CO ($4 \to 3$) map overlayed on the high-resolution
CO ($1 \to 0$) map taken by Brooks et al (1998) with the MOPRA
telescope.  The MOPRA map region is indicated by the
boundaries of the halftone figure.

\noindent
Figure 6.  Velocity channel maps of the CO ($4 \to 3$) emission.
Contour levels are 10\%--90\% of 
$40 \, {\mathrm K \, km \, s^{-1}}$.

\noindent
Figure 7.  Velocity channel maps of the [C~{\rmsmall I}] emission.
Countour levels are 10\%--90\% of 
$12 \, {\mathrm K \, km \, s^{-1}}$.

\noindent
Figure 8. Composite spectra averaged for each region 1--6
as identified in Figure 3.  CO ($4 \to 3$) spectra are plotted as solid lines;
[C~{\rmsmall I}] spectra are plotted as dashed lines.

\noindent
Figure 9. Excitation temperature (in degrees Kelvin) derived from the
CO ($4 \to 3$) and CO ($1 \to 0$) integrated intensity map.  
The CO ($4 \to 3$)
emission is convolved to the same spatial and velocity
resolutions as the CO ($1 \to 0$) data.

\noindent
Figure 10. 
[C~{\rmsmall I}]
integrated intensity overlayed on the
IRAS 100 $\mu$m emission.
The 100 $\mu$m emission is
in units of ${\mathrm MJy \, sr^{-1}}$, 
and the CI contours are from 2 - 12 ${\mathrm K \, km \, s^{-1}}$
with a spacing of 2 ${\mathrm K \, km \, s^{-1}}$.

\noindent
Figure 11. 
CO ($4 \to 3$) 
integrated intensity overlayed on the
IRAS 12 $\mu$m emission.
The 12 $\mu$m emission is
in units of ${\mathrm MJy \, sr^{-1}}$,
and the CO ($4 \to 3$) contours
are from 10\% - 90\% of 80 ${\mathrm K \, km \, s^{-1}}$.

\noindent
Figure 12. CO ($4 \to 3$) integrated intensity overlayed on the
IRAS 100 $\mu$m emission.  
The 100 $\mu$m emission is
in units of ${\mathrm MJy \, sr^{-1}}$,
and the CO ($4 \to 3$) contours
are from 10\% - 90\% of 80 ${\mathrm K \, km \, s^{-1}}$.

\noindent
Figure 13. 
[C~{\rmsmall I}]
versus CO ($4 \to 3$) integrated intensity
scatter plot.

\noindent
Figure 14. 
[C~{\rmsmall I}]
versus CO ($4 \to 3$) ratio from the standard PDR theory calculation for
different UV intensity and gas volumn density (courtesy Mark Wolfire).

\noindent
Figure 15. Size-linewidth relation for the clumps
in the Carina molecular cloud complex, derived from
the CO ($4 \to 3$) data cube.

\vfill
\eject

\section*{References}

\noindent
Balbus, S.A., \& Cowie, L.L. 1985, ApJ, 297, 61

\noindent
Brooks, K.J., Whiteoak, J.B., \& Storey, J.W. 1998,
PASA, 15, 202

\noindent
Carlstrom, J.E. \& Zmuidzinas, J. 1995, in
{\it Reviews of Radio Science 1993-1995}, ed. Stone, W.R. 
(Oxford Univ. Press:Oxford)

\noindent
Clari\'a, A. A. 1977, A\&AS, 27, 145

\noindent
Cox, P. 1995, in RevMexAA, 2, 105

\noindent
Dame T.M., Elmegreen B., Cohen, R. \& Thaddeus, P. 1986,
ApJ, 305, 892

\noindent
Das, M. \& Jog, C.J. 1995, ApJ, 451, 167

\noindent
Dickman, R.L., Snell, R.L., \& Schloerb, F.P. 1986, ApJ, 309, 326.

\noindent
Duschl, W.J., Hofmann, K.H., Rigaut, F., \& Weigelt G. 
1995, RevMexAASC, 2, 17

\noindent
Elmegreen, B.G. 1979, ApJ, 231, 372

\noindent
Elmegreen, B.G. 1987, 312, 626

\noindent
Elmegreen, B. 1999, in Proceedings of Star Formation 1999, Ed T. Nakamoto
(Nobeyama: Nobeyama Radio Observatory), p.3

\noindent
Feinstein, A. 1995, RevMexAASC, 2, 57

\noindent
Fleck, R.C. 1981, ApJL, 246, L151

\noindent
Feinstein, A., \& Marraco, H. G. 1980, PASP, 92, 266
\noindent
Flower, D.R., Le Bourlot, J., Pineau des F\^{o}rets, G., \&
Roueff, E. 1994, A\&A, 282, 225

\noindent
Ghosh, S.K., Ivengar, K.V.K., Rengarajan, T.N.
Tandon, S.N., Verma, R.P., \& R.R. Daniel
1988, ApJ, 330, 928

\noindent
Grabelsky, D.A. Cohen, R.S., Bronfman, L., 
\& Thaddeus P. 1988, ApJ, 331, 181

\noindent
Heyer, M. 1998, ApJ, 502, 265

\noindent
Hollenbach, D.J., Takahashi, T. \& Tielens, A.G.G.M. 1991, ApJ, 377, 192

\noindent
Hollebach, D.J. \& Tielens, A.G.G.M. 1999, Rev. Modern Phys.,
vol.71, no.1, 173 

\noindent
Huang, M. et al. 1999, ApJ, 517, 282

\noindent
Kornreich, P. \& Scalo, J. 2000, ApJ, 531, 366

\noindent Kaufman, M.J., Wolfire, M.G., Hollenbach, D.J., \&
Luhman, M.L. 1999, astro-ph/9907255
 
\noindent
Keene, J., Blake, G.A., Phillips, T.G., Huggins, P.J., \&
Beichman, C.A. 1985, ApJ, 299, 967

\noindent
Keene, J., Lis, D.C., Phillips, T.G. \& Schilke, P. 1997,
in ``Molecules in Astrophysics: Probes and Processes'' Proc. IAU Symp. 178,
ed. E.F. van Dishoeck (Dordrecht: Kluwer)

\noindent
Lada, E.A., Bally, J. \& Stark, A.A. 1991, ApJ, 368, 432.

\noindent
Lane, A.P., \& Stark, A.A. 1996, Antarctic J. of the U.S., 30 (5), 377

\noindent 
LaRosa, T.N., Shore, S.N., Magnani, L. 1999, ApJ (in press) {\tt astro-ph/9809245}

\noindent 
Larson, R.B. 1981, MNRAS, 194, 809 

\noindent
Le Bourlot, J., Pineau des F\^{o}rets, G., Roueff, E., \&
Schike, P. 1993, ApJ, 416, L87

\noindent
Lee, Y., Snell, R.L., \& Dickman, R.L. 1996, ApJ, 472, 275

\noindent
Lonsdale, G., Helou, G., Good, J.C.,\& Rice, W. 1985,
catalogued Galaxies and Quasars Observed in the IRAS Survey
(Pasadena: Jet Propulsion Laboratory)

\noindent
Magnani, L., LaRosa, T.N. \& Shore, S. 1993, ApJ, 402, 226

\noindent
Meixner, M. \& Tielens, A.G.G.M. 1993, ApJ, 405, 216

\noindent
Meixner, M. \& Tielens, A.G.G.M. 1995, ApJ, 446, 907

\noindent
Melnik A.M., \& Efremov Yu.N. 1995,
Pis'ma Astron. Zh., 21, 13>

\noindent
Myers, P.C. 1983, ApJ, 270, 105

\noindent
Phillips, T., \& Huggins, P. J. 1981, ApJ, 251, 533

\noindent
Pineau des F\^{o}rets, G., Roueff, E., \& Flower, D.R. 1992, MNRAS, 258, 45

\noindent
Plume, R. 1995, Ph.D. Dissertation, U. Texas at Austin

\noindent
Roberts, W. W. 1969, ApJ, 158, 23

\noindent
Schieder, R., Tolls, V., \& Winnewisser, G. 1989, Exp. Astro., 1, 101

\noindent
Schroder, K., Staemmler, V., Smith, M.D., Flower, D.R.,
\& Jacquet, R. 1991, J. Phys. B., 24, 2487

\noindent
Spaans, M. 1996, A\&A, 307, 271

\noindent
Stark, A.A. 1979, Ph. D. Dissertation, Princeton University

\noindent
Stark, A.A., Chamberlin, R.A., Cheng, J., Ingalls, J.G., \&
Wright, G. 1997, Rev. Sci. Inst., 68 (5), 2200

\noindent
Stutzki, J., Stacey, G.J., Genzel, R., Harris, A. I., Jaffe, D.T. \& Lugten, J.B. 1988,
ApJ, 332, 379

\noindent
Solomon, P.M., Rivolo, A.R., Barrett, J., \& Yahil, A. 1987,
ApJ, 319, 730

\noindent
Tielens, A.G.G.M \& Hollebach, D.J. 1985, ApJ, 291, 722

\noindent
Turner , D.G., Grieve, G.R., Herbst, W., \& Harris. W.E. 1980,
A.J. 85, 1193

\noindent
Ulich, B.\ L., \& Haas, R.\ W. 1976, Ap J Suppl. Ser., 30, 247.

\noindent
van Dishoeck, E.F., \& Black, J.H. 1988, ApJ, 334, 771

\noindent
Viotti, R. 1995, RevMexAASC, 2, 57

\noindent
Viscuso, P.J. \& Chernoff, D.F. 1988, ApJ, 327, 364

\noindent
von Weizsacker, C.F. 1951, ApJ, 114, 165

\noindent
Walborn, N.R. 1995, RevMexAASC, 2, 51

\noindent
Zhang X. 1992, Ph.D. Dissertation, Univ. of California, Berkeley

\noindent
Zhang, X. 1996, ApJ, 457, 125

\noindent
Zhang, X. 1998,  ApJ, 499, 93
 
\noindent
Zhang, X. 1999,  ApJ, 518, 613
 
\noindent
Zhang, X. 2000,  submitted to the ApJL.

\vfill
\eject

\vfill
\eject

\end{document}